 \documentclass[twocolumn]{aastex61}
\newcommand{\uhuru}{{\it Uhuru}}
\newcommand{\chandra}{{\it Chandra}}

\newcommand{\asca}{{\it ASCA}}

\newcommand{\suzaku}{{\it Suzaku}}

\newcommand{\swift}{{\it Swift}}

\newcommand{\nustar}{\textit{NuSTAR}}

\newcommand{\ms}{$M_{\odot}$}

\newcommand{\fluxcgs}{ergs~s$^{-1}$~cm$^{-2}$~\rm}

\newcommand{\chisq}{$\chi^{2}_{\nu}$}

\received{November 10th 2017}
\revised{}
\accepted{December 15th 2017}

\shorttitle{EX Hya accretion column}
\shortauthors{Luna et al.}
\usepackage{twoopt,amssymb,wasysym,textcomp}

\begin{document}

\title{Constraining the accretion geometry of the intermediate polar EX~Hya using \nustar, \swift~ and \chandra~ observations.}

\correspondingauthor{Gerardo Juan Manuel Luna}
\email{gjmluna@iafe.uba.ar}

\author{G. J. M. Luna}
\affiliation{CONICET-Universidad de Buenos Aires, Instituto de Astronom\'ia y F\'isica del Espacio, (IAFE), Av. Inte. G\"uiraldes 2620, C1428ZAA, Buenos Aires, Argentina}
\affiliation{Universidad de Buenos Aires, Facultad de Ciencias Exactas y Naturales, Buenos Aires, Argentina}
\affiliation{Universidad Nacional Arturo Jauretche, Av. Calchaqu\'i 6200, F. Varela, Buenos Aires, Argentina}
\affiliation{INAF - Osservatorio di Padova, vicolo dell’ Osservatorio 5, I-35122 Padova, Italy}

\author{K. Mukai}
\affiliation{CRESST and X-ray Astrophysics Laboratory, NASA Goddard Space Flight Center, Greenbelt, MD 20771, USA}
\affiliation{Department of Physics, University of Maryland, Baltimore County, 1000 Hilltop Circle, Baltimore, MD 21250, USA}

\author{M. Orio}
\affiliation{INAF - Osservatorio di Padova, vicolo dell’ Osservatorio 5, I-35122 Padova, Italy}
\affiliation{Department of Astronomy, University of Wisconsin, 475 N. Charter Str., Madison, WI 53704, USA}

\author{P. Zemko}
\affiliation{Department of Physics and Astronomy, Universit`a di Padova, vicolo dell’ Osservatorio 3, I-35122 Padova, Italy}



\begin{abstract}
In magnetically accreting white dwarf, the height above the white dwarf surface where the standing shock is formed is intimately related with the accretion rate and the white dwarf mass. However, it is difficult to measure. We obtained new data with \nustar\ and \swift\, that together with archival \chandra\ data
allow us to constrain the height of the shock 
in the intermediate polar EX~Hya. 
We conclude that the shock has to form at least at a distance of about one white dwarf radius from the surface in order to explain the weak Fe K$\alpha$ 6.4 keV line, the absence of a reflection hump in the high energy continuum and the energy dependence of the white dwarf spin pulsed fraction. Additionally, the \nustar\ data allowed us to measure the true, uncontaminated hard X-ray (12-40 keV) flux, whose measurement was contaminated by the nearby galaxy cluster Abell~3528 in non-imaging X-ray instruments.

\end{abstract}




\section{Introduction} 
\label{sec:intro}


In magnetic cataclysmic variables (CVs) the primary is a highly magnetized (B$\gtrsim$ 10$^{6}$ G) white dwarf (WD) whose field controls the accretion flow close to the WD, leading to a shock and accretion columns that radiate chiefly in X-rays. 
The shock temperature kT$_{sh}$ is determined by the pre-shock velocity and is of order 10--50 keV in magnetic CVs. The post-shock plasma must further decelerate and cool before it can settle onto the WD. Thus the height of the shock (h$_{sh}$) is determined by equating the plasma cooling time with the remaining travel time from the shock front to the WD surface \citep{aizu}. Since the X-ray cooling time is inversely proportional to the density of the post-shock plasma, h$_{sh}$ is small if the accretion rate per unit area (or specific accretion rate) is high. If h$_{sh}$ is a small fraction of the white dwarf radius (R$_{WD}$), and if the accretion flow can be considered to be free-falling from infinity, then kT$_{sh}$ is an immediate indicator of the white dwarf mass (M$_{WD}$). This works well for most intermediate polars (IPs). However, if the specific accretion rate is low, h$_{sh}$ may not be negligible. This would reduce the free-fall velocity above the shock, and hence kT$_{sh}$. Also, if accretion is from a truncated disk with a small inner radius R$_{in}$, the pre-shock velocity is set by free-fall condition from R$_{in}$, requiring a different correction \citep{sule05,luna15}.

The subject of this letter, \object{EX~Hya}, a unique IP that has raised several important, and still unresolved questions.
One is the very nature of its accretion flow. A standard, Keplerian, partial disk cannot be present in this system if the WD is in spin equilibrium, because R$_{in}$ (given the long spin period, 67 min, relative to the orbital period, 98 min) would be so large as to violate the physical condition for the formation of a disk \citep{kinglasota}. Either the WD is far out of equilibrium, or \object{EX~Hya} possesses a diamagnetic blob/ring type structure between the magnetosphere and L1 \citep{king,norton}.

The other major unresolved question is why the X-ray spectrum of \object{EX~Hya} is so soft. The combination of the partial eclipse and optical spectroscopy has led to an estimate of M$_{WD}$=0.78$\pm$0.03 M$_\odot$ \citep{echevarria}, implying kT$_{sh} \sim$ 35 keV in the Aizu picture, while X-ray measurements are consistently below $\sim$20 keV (see, e.g., \citealt{luna15}). This can be resolved by either having a small R$_{in}$ or a large h$_{sh}$.



In EX~Hya, h$_{sh}$ has been inferred through indirect arguments and some of them yielded opposite answers. \citet{allan98} studied the spin modulation and the partial eclipse in the \asca\ data, and argued for a tall (h$_{sh} \sim 1$ R$_{WD}$) shock as the explanation for the spin modulation, and also large (R$_{in}$ $>$ 6.1 R$_{WD}$) inner radius of the disk. A smaller R$_{in}$ would result in the accretion disk blocking our view of the lower pole. Other arguments supporting that R$_{in} >$ a few R$_{WD}$ and thus that h$_{sh}$ is a non-negligible fraction of R$_{WD}$ are: $i$) the equilibrium spin period is expected to be close to the Keplerian period at R$_{in}$; for small values of R$_{in}$, the spin period would be smaller than the observed 67 m. $ii$) \citet{hellier87} analyzed an extensive set of optical spectra and found three components in the line profile: a narrow $S$-wave component; a double-peaked component; a broad, spin-modulated component. Measuring the width at spin phase 0.5 allowed them to put an upper limit on the high velocity extent of the double–peaked, presumably accretion disk, concluding that for a 0.78 M$_{WD}$, R$_{in}$ is about 10 R$_{WD}$.

On the other hand, \citet{2011MNRAS.411.1317R} and \citet{2014MNRAS.442.1123S} modeled the break frequency in the power spectrum of stochastic variability and proposed R$_{in}$=2.7 R$_{WD}$. \citet{belle03}, \citet{sule16} and \citet{echevarria} also derived small R$_{in}$ using other methods. 
This small R$_{in}$ would reduce kT$_{sh}$: this was the solution preferred by \citet{luna15}, who analyzed high quality, half a megasecond, \chandra\ HETG data on EX~Hya and found a number of discrepancies with respect to the standard models of X-ray emission. A tall shock would imply that gravity adds heat to the cooling flow region and the magnetic field geometry adds heat by magnetic pressure. Both mechanisms should modify the emission measure distribution of the cooling plasma, increasing both the flux of H and He-like lines. Such models did not match the \chandra/HETG observations.



An observational determination of h$_{sh}$ is essential to decide 
which picture is correct.
Because of the accretion geometry, we expect the X-rays emitted in the post-shock region to be reflected back into our line of sight by the WD surface, producing a detectable Compton hump at energies above 10 keV \citep[e.g.,][]{hayashipp}. The detection of the reflection feature in three IPs has been possible only recently with \nustar\ (Mukai et al. 2015).
Constraining the presence of a reflection component in the hard X-ray spectrum of EX~Hya would allow to determine h$_{sh}$.
A tall shock implies little-to-no reflection with a weak Fe K$\alpha$ 6.4 keV fluorescence line and small or non-existent spin modulation above $∼$ a few keV, since photoelectric absorption on the order of 10$^{22}$ cm$^{-2}$ cannot affect the light curves at these energies (the measured N$_{H}$ is even lower) and the shock region would not be hidden by the WD body. On the other hand, a negligible shock height would imply a strong reflection amplitude, a strong Fe K$\alpha$ line, while the spin modulation would be almost entirely due to absorption and thus the pulsed fraction of the light curves should be a strong function of the energy. Somewhere in the middle, a shock height of a non-negligible fraction of the WD radius ($\sim$0.1-0.5 R$_{WD}$ ) would imply a moderate reflection amplitude (less than 1), potentially detectable with \nustar. The modulation of the low energy X-rays (E$\sim$5-10 keV) would be mostly due to occultation of the accretion column by the body of the WD while the expected modulation at higher energies would be entirely due to occultation, so the pulsed fraction at high energies should not be a function of energy. To perform this test, we have observed \object{EX~Hya} with \nustar.

There is an additional reason why it is important to perform a
\nustar\ observation of \object{EX~Hya}. While it was proposed to be
the counterpart of a \uhuru\ source from the early days of X-ray
astronomy \citep{warner}, and it is indeed the brightest source in
the traditional (0.5--10 keV) X-ray band among all CVs, it is not
the only bright X-ray source in this region of the sky. The interacting
cluster of galaxies, \object{Abell~3528}, consists of two X-ray bright
subclusters \citep{gastaldello}, located $\sim$29 arcmin from \object{EX~Hya}.
Therefore, the possibility of contamination must be kept in mind in
interpreting any non-imaging X-ray observations of \object{EX~Hya}.
This means, in part, that there have been no reliable observations of
\object{EX~Hya} above 10 keV until now.

In this letter, we present new contemporaneous \nustar\ and \swift\ observations of EX~Hya, and new analysis of archival \chandra~\rm observation. We present conclusive arguments for an absence of reflection in the X-ray spectrum, implying that the height of the shock must be an appreciable fraction of the size of the WD. In Section \ref{sec:obs} we detail the reduction of \nustar, \swift~\rm and \chandra~\rm data while Section \ref{sec:results} presents the results from the spectral and timing analysis. Finally, Section \ref{sec:disc} presents a discussion about the implications of the non-detection of reflection for the structure of the accretion column. 



\section{Observations. \label{sec:obs}}

We observed \object{EX~Hya} with \nustar~\rm on 2016-06-05 for 24.8 ks. The data were reduced using the \nustar\ Data Analysis Software as part of HEASOFT 6.21 and filtered using standard filters given that the observation was not affected by unnormal solar activity. Using the tool \texttt{nuproducts} we extracted source spectra, baricenter-corrected source event files and light curves from a circular region centered on the SIMBAD coordinates, $\alpha$=12h 52m 18.5s, $\delta$=-29$^{\circ}$ 16$^{\prime}$ 16$^{\prime\prime}$ in the FPMA chip and $\alpha$=12h 52m 23.8s, $\delta$=-29$^{\circ}$ 14$^{\prime}$ 55.1$^{\prime\prime}$ in the FPMB chip, with a 30$^{\prime\prime}$ radius. For the background, we choose an annular region with inner and outer radii of 110$^{\prime\prime}$ and 220$^{\prime\prime}$, respectively and centered on the respective source coordinates.

A \swift\ observation was obtained almost simultaneously with \nustar, with 1.8 ks exposure time. We extracted source X-ray spectra from a circular region with a radius of 20 pixels centered on the SIMBAD coordinates. We extracted background events from an annular region with inner and outer radii of 25 and 40 pixels, respectively. 
We built the ancillary matrix (ARF) using the tool \texttt{xrtmkarf} and used the \texttt{swxpc0to12s6\_20130101v014.rmf} response matrix provided by the \swift\ calibration team.

EX~Hya was observed with \chandra\ using the ACIS-S/HETG combination for 496 ks and the spectral analysis has been already described in Luna et al. (2010, 2015). For the present study, the events arrival times were barycentrically corrected using the \texttt{axbary} script and filtered to extract event arrival times from the source in the energy regions of interest (strongest emission lines), 
selecting only HEG and MEG $\pm$1 orders. 


\section{Results. \label{sec:results}}

\subsection{Spectral model}

We first modeled the continuum of \nustar\ spectrum excluding the Fe and Ni lines region (5--9 keV) and because both internal and interstellar absorption are known to be small in EX~Hya ($\lesssim$10$^{21}$ cm$^{-2}$), we only used a single-temperature model modified by reflection. First we used solar abundances for the reflecting plasma\footnote{Throughout our spectral analysis, the derived elemental abundances refer to the solar abundances from \citet{abund}}. This model is statistically acceptable, with \chisq=1.06/187 d.o.f, a temperature of $kT$=9.9$\pm$0.6 keV and an unconstrained reflection amplitude of $<$ 0.29 (see Table \ref{tab:model}). 
However, most X-ray data indicate sub-solar Fe abundances \citep[e.g.][]{allan98,luna15}, with $\approx$ 60\% the solar value. Fixing the Fe abundance to 0.60 yielded a similarly acceptable fit with \chisq=1.06/187 d.o.f, $kT$=9.9$\pm$0.6 and equally unconstrained reflection amplitude. 

We also tested models where we included back the Fe and Ni lines regions and used a variable abundance, multi-temperature plasma (\texttt{vmcflow}) plus a gaussian line to account for the presence of the Fe K$\alpha$ 6.4 keV fluorescence line. First, fixing the Fe abundance to 0.60 (Ni abundances is tied to Fe in our fits) yielded a maximum temperature $kT_{max}$=13.4$\pm$0.8 keV and a reflection amplitude of 1.24$\pm$0.33, with \chisq=1.19/347 d.o.f. This same model without reflection yielded \chisq=1.33/348 d.o.f and $kT$=17.7$\pm$0.4 keV.
If we allow the Fe abundance to vary, a multi-temperature model without reflection yielded $kT_{max}$=18.2$\pm$0.5 keV and Fe abundance of 0.87$\pm$0.05 Fe$_{\odot}$ with \chisq=1.06/347 d.o.f. 
Including reflection, the fit, with \chisq=1.06/346 d.o.f, yielded a $kT_{max}$=18.3$_{-0.6}^{+0.4}$ keV, Fe/Fe$_{\odot}$=0.87$\pm$0.04 and reflection amplitude $\lesssim$ 0.08. 

Because the reflection amplitude depends on the Fe abundance, and the \nustar\ data with their low spectral resolution are well suited to fit the continuum but not the spectral lines, we also included in the fit of our multi-temperature model the \chandra\ HETG and \swift\ data. For the \chandra\ data, we only used the 3.0--8.0 keV energy range; lower energies are dominated by soft emission lines that are not addequately described by isobaric, multi-temperature spectral models 
\citep{mukai03,luna15}.
We let the \nustar\ data to drive the fit of the reflection and cooling flow temperature while the \chandra\ data drove the Fe abundance. The fit led to a \chisq=1.17/1900 d.o.f, $kT$=19.7 keV, Fe abundance of Fe/Fe$_{\odot}$=0.88 and an unconstrained reflection amplitude of $\lesssim$0.15.
Once we consider the same model without reflection, we have $kT$=19.7 keV, Fe/Fe$_{\odot}$=0.88
and \chisq=1.17/1899 d.o.f. No differences with the previous model owing to the undetectable reflection. The strength of the Compton hump due to reflection will be small if we see the reflection surface edge-on, which will imply that the parameter $\cos(\mu)$ is closer to zero. We tried models with low $\cos(\mu)$ of 0.1 and found that the shock temperature and the negligible reflection amplitude are insensitive to the value of $\cos(\mu)$, reinforcing our contention that the reflection component is weak or absent. 


\begin{figure}
\includegraphics[scale=0.30]{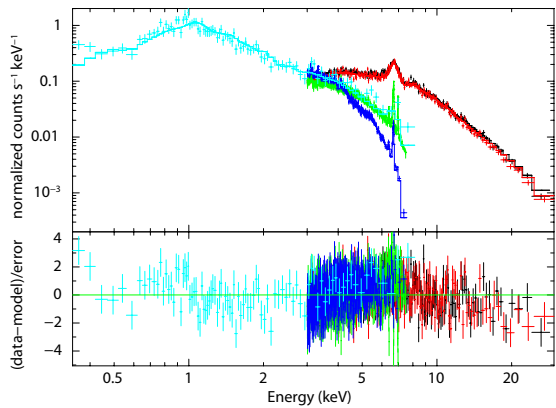}
\caption{\nustar\ (black and red) + \swift\ (light blue) + \chandra\ (MEG, blue and HEG, green) \object{EX~Hya} spectra modeled with a multi-temperature, variable abundances, isobaric cooling flow model plus a fluorescence Fe K$\alpha$ emission line (\texttt{constant$\times$TBabs$\times$reflect$\times$(vmkcflow+gauss)}). Lower panel shows the fit residuals in units of $\chi^{2}$.  \label{fig:spec}}
\end{figure}

\begin{deluxetable*}{cccccc}
\tablecaption{Spectral fit results \label{tab:model}}
\tablecolumns{6}
\tablewidth{0pt}
\tablehead{
\colhead{Data} &
\colhead{Model} &
\colhead{\chisq/dof} & \colhead{kT$_{max}$ [keV]} & \colhead{Fe/Fe$_{\odot}$} &
\colhead{Amplitude} }
\startdata
\nustar & brems$\times$ref & 1.27/212 & 9.9$\pm$0.6 & 1 & $\lesssim$0.29 \\ 
\nustar & brems$\times$ref & 1.27/212 & 9.9$\pm$0.6 & 0.6 & $\lesssim$0.29 \\ 
\nustar & vmcflow$\times$ref & 1.19/347&13.4$\pm$0.8 & 0.6 & 1.24$\pm$0.33 \\ 
\nustar & vmcflow & 1.33/348&17.7$\pm$0.4 & 0.6 & \nodata \\ 
\nustar & vmcflow & 1.06/347 & 18.2$\pm$0.5 & 0.87$\pm$0.05 & \nodata \\ 
\nustar & vmcflow$\times$ref & 1.06/346 & 18.3$^{+0.4}_{-0.6}$ & 0.87$\pm$0.04 & $\lesssim$0.08 \\ 
\nustar\ + \swift\ + \chandra &vmcflow$\times$ref&1.17/1900 & 19.7$\pm$0.4& 0.88$\pm$0.02  & $\lesssim$0.15\\ 
\nustar\ + \swift\ + \chandra & vmcflow & 1.17/1989 & 19.7$\pm$0.4 & 0.88$\pm$0.02 & \nodata \\ 
\enddata

\end{deluxetable*}

\subsection{The un-contaminated hard X-ray flux.}


The best-fit spectral model from \nustar\ + \swift\ + \chandra\ data yielded a 12--40 keV flux 3.3$\times$10$^{-11}$ \fluxcgs and when this model is applied to the \suzaku/HXD data, it yielded a 12--40 keV flux 3.9$\times$10$^{-11}$ \fluxcgs. \citet{yuasa} quoted a 12--40 keV flux of 3.56$\times$10$^{-11}$ \fluxcgs from their modeling of \suzaku~\rm data. The difference can be attributed to the high absorption column quoted by \citet{yuasa} 
(see their Table 2). We conclude that the real hard energy flux has been contaminated by $\approx$20\%. Our model 
also yielded a mass accretion rate $\dot{M}$=1.89$\times$10$^{-11}$ \ms yr$^{-1}$. Note that while modeling the \chandra\ spectrum, \citet{luna15} used $\dot{M}$ of 1.74 $\times$10$^{-11}$ \ms yr$^{-1}$ while \citet{isakova} used $\dot{M}$ of 4.75$\times$10$^{-11}$ \ms yr$^{-1}$ in their numerical simulations of the accretion flow.

\subsection{Timing analysis: power spectrum and pulsed fraction.}
\label{sec:time}

In order to 
study 
dependence of the WD spin pulsed fraction with energy
from the photons arrival times we calculated the $Z_{1}^{2}$ (Rayleigh) statistic \citep{buccheri} as a function of frequency 
in the range 0.00022 Hz $< f <$ 0.0003 Hz \citep[the WD spin period is 67.02696576 min or 0.00024865614 Hz;][]{mauche09} with a step $\Delta f$=1.0/($T\times q$), where $T$ is the exposure time 
and $q$ is the oversampling factor, which we took equal to 1000. As the \swift\ data do not cover a single spin period, we did not include them in this analysis.
The value of $Z_{1}^{2}$ needed to detect a pulsation with a probability $P$=2.699$\times$10$^{-3}$ (3-$\sigma$ detection) is $Z_{1}^{2} > 2\ln{(\frac{T\Delta f}{P})}$. 
If the peak in the power spectrum 
is due to nearly sinusoidal modulations, the pulsed fraction is $p=p_{obs}(N_{S}+N_{B})N_{S}^{-1}$, where $N_{S}$ and $N_{B}$ are the number of source and background counts and $p_{obs}$ is the observed pulsed fraction uncorrected by background. However, as EX~Hya is a very bright X-ray source, the background contribution is negligible and the pulsed fraction can be expressed as $p \simeq \sqrt{2Z_{1}^{2}} N_{S}^{-1/2} \pm 2N_{S}^{-1/2}$. 

\begin{figure}
\includegraphics[scale=0.5]{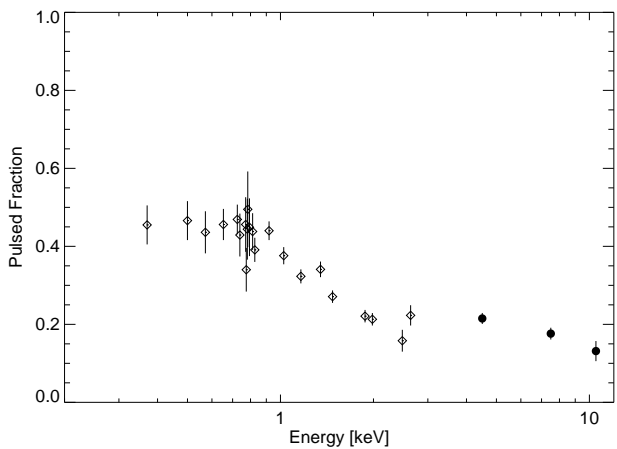}
\caption{Pulsed fraction versus energy derived from the \nustar\ observation in the energy intervals of 3--6, 6--9 and 9--12 keV (filled circles). Open diamonds show the pulsed fraction derived from the emission lines in the \chandra\ HETG spectrum. The lines are those listed in Table 1 in \citet{luna15}.}
\label{fig:pfrac}
\end{figure}

In Figure \ref{fig:pfrac} we plot the pulsed fraction, of those spin periods detected with $>$3$\sigma$ significance, in the energies of the strongest emission lines as observed in the \chandra/HETG spectra \citep[and whose fluxes were measured in][]{luna15} and the broad energy bands of 3--6, 6--9 and 9--12 keV in the \nustar\ data. Modulation at the spin period is detected up to energies of less than $\sim$ 12 keV in the \nustar\ data.
We found that there is a dependence of the pulsed fraction with energy. For energies of less than about 1 keV, the pulsed fraction seems to be constant with energy.
On the other hand, for energies greater than 1 keV, the pulsed fraction decays with energy.
The low absorption cannot be responsible for the modulation at these energies
The origin of this effect remains a mistery. The non-detection of pulsation at energies greater than $\sim$ 12 keV indicates that the height of the shock is at least greater than 1R$_{WD}$ but not big enough to get the lower pole occulted by the inner region of the accretion disk.

\begin{figure}[ht!]
\includegraphics[scale=0.5]{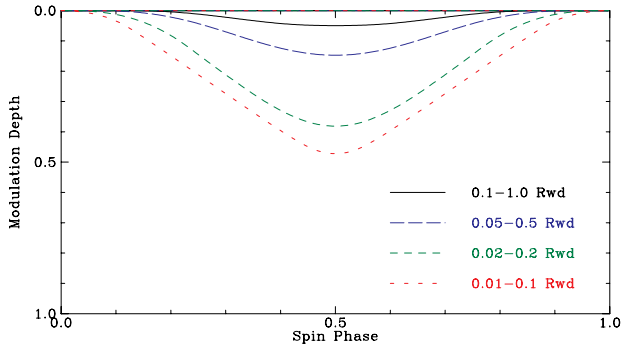}
\caption{Modulation depth for different heights in the post-shock region. }
\label{fig:simlcs}
\end{figure}

We can qualitatively understand the energy dependence of spin modulation amplitude as a consequence of the height dependence of both the physical condition and the visibility. In the post-shock region, the temperature is at its heighest near the shock and declines towards the white dwarf surface. The density, on the other hand, is at its lowest near the shock and increase near the surface. Continuum photons of energy $E$ originate from regions where kT$\gtrsim\ E$; line photons originate from a limited range of temperatures (e.g., \ion{Ne}{10} lines require temperature of order 0.54 keV; neither regions that are too cool or too hot contribute significantly). Thus photon energy plotted along the X-axis of Figure \ref{fig:pfrac} is a proxy for the origin of these photons within the post-shock region. If the shock is tall, of order h$_{sh}\sim$1 R$_{WD}$, then the highest temperature continuum will escape self-occultation almost completely. The lowest energy lines are emitted only near (but still above) the white dwarf surface. When we view the poles at right angle, both poles are visible; half a spin cycle later, most of the lower pole is behind the body of the white dwarf, with a small residual that depends on the geometrical extent of the accretion footpoint. In Figure \ref{fig:simlcs}, we show the result of a proof-of-concept simulation, in which light curves for uniform emission regions with limited range of h$_{sh}$ have been simulated. Following the two-stage process explained in \citet{mukai99}, the arc-shaped accretion footpoints were calcuated assuming a rigid magnetic dipole with a magnetic colatitude of 5 degrees, accreting uniformly from a transition region at the inner disk edge at 9-10 R$_{WD}$ (each pole accreting from an 180 degrees azimuth).
quantitative model including the location and shape of the threading region, and the resulting shock structure, is beyond the scope of this paper.


\section{Discussion and conclusions. \label{sec:disc}}

The non-detection of a reflection component in the hard X-ray spectrum taken with \nustar, the small FeK$\alpha$ equivalent width of $\sim$25 eV and the absence of spin modulation for energies greater than $\sim$ 12 keV implies that h$_{sh}$ must be an appreciable fraction of the WD radius. This can be used to distinguish between the two possible origins (large h$_{sh}$ and small R$_{in}$) of low kT$_{sh}$ in \object{EX~Hya}. 

\begin{figure}[!ht]
\includegraphics[scale=0.5]{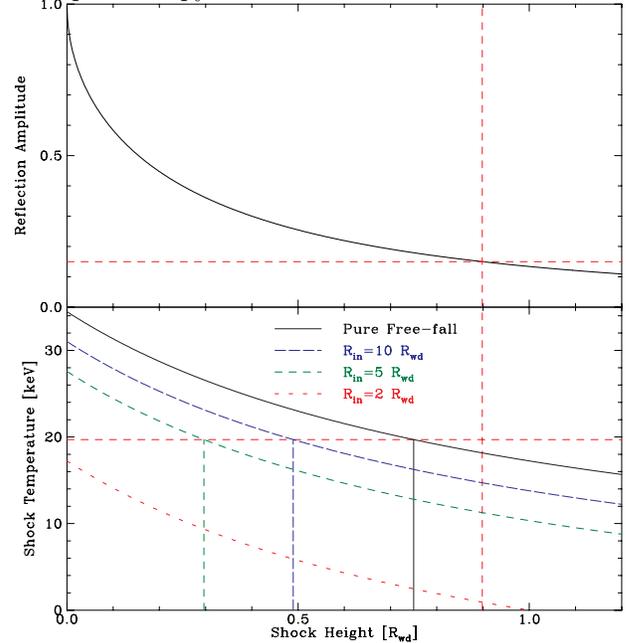}
\caption{{\em Top}: The expected reflection amplitud as a function of h$_{sh}$. {\em Bottom} kT$_{sh}$ as a function of h$_{sh}$, assuming 4 different values of R$_{in}$, including pure free-fall, and a WD mass of 0.78 \ms. The horizontal line at 19.7 keV, marks the kT$_{sh}$ that we measured (see Section \ref{sec:results}.  \label{fig:ktmap}}
\end{figure}

In Figure \ref{fig:ktmap}, the top panel shows the reflection amplitude as a function of h$_{sh}$ for a point-like emitting region.
The upper limit on the reflection amplitude of $\lesssim$0.15 (see Table \ref{tab:model}) implies h$_{sh}\lesssim$0.9$\times$R$_{WD}$. Also, the EW of FeK$\alpha$ implies a reflection amplitude of about 0.15 in the model presented by \citet{george}. The bottom panel in Fig. \ref{fig:ktmap} shows a set of four curves of kT$_{sh}$ as a function of h$_{sh}$, for 4 different assumed values of R$_{in}$, for a 0.78 \ms white dwarf, with a horizontal line at 19.7 keV, the kT$_{sh}$ that we measured. 
Accounting for the error bar in the kT$_{sh}$, the reflection amplitude upper limit and our simplified reflection model, the data presented here are compatible with a large R$_{in}$.Moreover, given the lack of strong reflection signature, a solution of R$_{in}$=2$\times$R$_{WD}$, h$_{shock}$=0 is clearly untenable for EX~Hya. 


The infered large R$_{in}$ is still much smaller than the co-rotation radius for a WD rotating with a 67 min period. In the diamagnetic blob scenario proposed by \citet{king} and \citet{norton}, which implies a perturbed accretion flow that deviates from purely Keplerian velocity structure, it is possible that the WD is in spin equilibrium. Note that, the magnitude of the spin-up in EX~Hya does not stand out among all IPs, most of which are presumably in spin equilibrium. Moreover, in this scenario, the very premise of the model of stochastic variability used by \citet{2011MNRAS.411.1317R} and \citet{2014MNRAS.442.1123S}), as well as other determinations of R$_{in}$ based on strictly Keplerian flows, are suspect. Alternatively, EX~Hya may possess a purely Keplerian accretion disk with a large R$_{in}$: in this case, the WD is far out of equilibrium, and we have no concrete explanation for the disagreement with the break frequency method. One remaining problem that still needs to be explained is the detailed soft X-ray emission line spectroscopy in the \chandra\ HETG data.


\acknowledgments

This research has made use of data obtained with the \nustar\ mission, a project led by the California Institute of Technology (Caltech), managed by the Jet Propulsion Laboratory (JPL) and funded by NASA. M. Orio acknowledges support from NASA grant NNX17AB76G.  We acknowledge the \swift\ team for planning this observation. 
GJML is a member of the CIC-CONICET (Argentina) and acknowledges support from grants PIP-Conicet/2011 \#D4598, ANPCYT-PICT 0478/14. 
%

\vspace{5mm}
\facilities{\nustar,\swift\ and \chandra}

\end{document}